\documentclass{vgtc}                          

\ifpdf
  \pdfoutput=1\relax                   
  \usepackage{graphicx}                
  \DeclareGraphicsExtensions{.pdf,.png,.jpg,.jpeg} 
\else
  \usepackage{graphicx}                
  \DeclareGraphicsExtensions{.eps}     
\fi%

\graphicspath{{figures/}{pictures/}{images/}{./}} 

\usepackage{microtype}                 
\PassOptionsToPackage{warn}{textcomp}  
\usepackage{textcomp}                  
\usepackage{mathptmx}                  
\usepackage{times}                     
\usepackage{cite}                      
\usepackage{tabu}                      
\usepackage{booktabs}                  

\onlineid{121}

\vgtccategory{Research}

\vgtcinsertpkg

\title{Visual Progression Analysis of Student Records Data}

\author{Mohammad Raji\thanks{e-mail: \{mahmadza, jduggan1, bdecoate, huangj, bvanderz\}@utk.edu} %
\and John Duggan
	\footnotemark[1]
\and Blaise DeCotes
	\footnotemark[1]
\and Jian Huang
	\footnotemark[1]
\and Bradley Vander Zanden
	\footnotemark[1]
}
\affiliation{\scriptsize University of Tennessee, Knoxville}

\abstract{
	University curriculum, both on a campus level and on a per-major level, are affected
in a complex way by many decisions of many administrators and faculty over time.
As universities across the United States share an urgency to
significantly improve student success and success retention, there is a pressing need
to better understand how the student population is progressing through the curriculum,
and how to provide better supporting infrastructure and refine the curriculum
for the purpose of improving student outcomes. This work has developed a visual
knowledge discovery system called eCamp that pulls together a variety of
population-scale data products, including student grades,
major descriptions, and graduation records.
These datasets were previously disconnected and only available to and maintained by
independent campus offices.
The framework models and analyzes the multi-level relationships
hidden within these data products, and visualizes the student flow patterns
through individual majors as well as through a hierarchy of majors.
These results support analytical tasks involving
student outcomes, student retention, and curriculum design.
It is shown how eCamp has revealed student progression information that was previously
unavailable.
} 

\begin{document}


\firstsection{Introduction}

\maketitle

\label{sec:introduction}
College is often known as the ``best 4 years of your life".
Not all students can graduate successfully, however, many may end up dropping out.
The attrition comes with significant pedagogical, economic, and societal costs.
The related concerns have been growing year over year, 
especially during the past decade in the United States~\cite{nchea2011,soares2013}.
Even though quite a few universities have invested substantially in programs
designed to increase student retention and success,
the success rate has not improved very much \cite{swail2014,tinto2010}.

In universities, there are sophisticated designs of how students are expected to progress 
through the curricula; and there are mechanisms put in place to support and foster
the process so that the intended outcomes are achieved for the students. The designs involve many decisions about student advising, curriculum design, overlaps between majors, and what choices students can make at different times about their college affiliation and degree programs.

Those design decisions are made cumulatively by many people involved,
sometimes based on theories,
sometimes based on convenience,
and sometimes based on subjective ``lore'' or
``feel'' that is derived from years of accumulated experience.
It is important for all people involved to have a clear and complete
view of the intrinsics in student progression processes.

Graph is a standard model to represent student 
progression processes. For example, course prerequisite relationships are often displayed as graphs. These graphs however, do not show how students actually progress, succeed or fail throughout their studies but rather give hints to what paths they should take. Graph models that show true progression are not always constructed and evaluated explicitly. In part, this is because campus administrative offices' view of the entire process is limited to their specific functions only.
In part, that's also because of not knowing how the student population would make choices when they are allowed choices. 
As the associate provost of the authors' university acknowledged - ``no one has the full picture".

In order to gain insights about student progression, student success, and student retention, a data science approach should examine the real world progression of
students, as opposed to the hypothetical progression codified in the course
catalog, graduation requirements, and advising guidelines.
This real world information exists in the form of population-scale student records data
that are available to university administrators,
but are typically spread amongst independent offices. 
The data includes course grades, student schedules, major
information, and university withdrawal rates. 

In this work, we have developed a visual analysis system, eCamp, to integrate these sources and model student progression patterns based on electronic records from about 145,000 students collected over the course of 16 years. At a high level suited for campus administrators and based on major similarities, eCamp constructs and visualizes a university-wide major-graph. The major-graph shows how a student population start together with general education courses as freshmen, and then diverge into more advanced and specialized parts of their curricula as they progress through their majors. eCamp also shows student drop-out occurrences in the overall major-graph to better understand which majors are struggling in student retention. At a more detailed level, suited for department heads and faculty, eCamp constructs a set of course-graphs for over 400 majors on campus. These course-graphs capture the curriculum structure of each discipline on a per-major basis by modeling course-course relationships. Through these graphs, administrators can see the real-world structures that are formed. Additionally, they can see how students progress or fail through these structures. 

When enabled by data, the Provost's office was interested in answering questions about how
the support infrastructure, which includes advising, tutoring, and supplemental assistance, was impacting
student success.
For example, it wondered how this infrastructure impacted
student progression through their major.
Many new students may wish to explore different majors to discover their true passion.
But this exploration process cannot be unlimited in time if
the students are to graduate in a reasonable amount of time.
To facilitate students' exploration and progress, it is important to know
when and where are the time points for each student to make critical decisions about their major path.
There are clearly gaps in the current way of providing advising
services (both on a campus level and on a department level).
But where are the gaps?

At the department level, questions from administrators revolved around whether
the curriculum is working as designed. For example, are the general education courses
preparing students for success in their major? Are gate keeping
courses serving their functions? As students progress through the major,
which courses play a central or peripheral role, and which courses are bottlenecks?
Are there critical time points where diversity and/or retention drop off?

Our visualizations help university staff to think of such questions. Additionally, it empowers them to formulate hypotheses about curriculum design and student outcomes that were not previously possible to
accurately articulate.


The rest of this paper 
is organized as follows. Section~\ref{sec:rel_work}
provides a discussion of related work. Section~\ref{sec:driving_app} describes
the motivation for this work and the needs of university administrators.
Section~\ref{sec:data_models} explains how student records are used to model
student behaviors. Section~\ref{sec:vis_analytics} explains the
visualizations created to facilitate interaction with the model and how these
visualizations meet the needs established in Section~\ref{sec:driving_app}.
Domain expert feedbacks are presented in Section~\ref{sec:feedback} and the paper is concluded in 
Section~\ref{sec:conclusion}.

\section{Related Work}
\label{sec:rel_work}

\subsection{Application Background}

Many tools for analyzing university databases exist \cite{attewell2011competing,clark2005negotiating,grann2014competency,kuh2011student,lutz2012instilling,mazza2005generation,mazza2004visualising,rueda2005visualizing}.
For example, DynMap visualizes student learning in a course
by offering an ability to visually inspect students' understanding and
performance in a concept map as well as display the overall
structure of the course topics and their dependencies
\cite{rueda2005visualizing}.
Similarly, CourseVis visualizes student tracking
data for a course management system, WebCT, for the purpose of analyzing
student progress within a course, targeting distance learning settings \cite{mazza2004visualising}.
CourseVis helped instructors see social,
cognitive, and behavioral aspects of their students through visualizations of
web log data from course management systems \cite{mazza2005generation}.
These systems tend to limit the scope of study to
student progress in an individual course.

Some works go beyond course boundaries and focus on course-course relationships. 
They can show student course progression as prescribed by the catalog \cite{siirtola2013interactive} or 
show student course grades based on the actual semesters in which students took
the courses \cite{gama2014visualizing}.

In contrast, eCamp focuses on visualizing student progression patterns
through the curricula of all majors in a university. Whereas the
largest granularity of data for previous work was typically a course or
smaller, the
smallest granularity of data is a course.
To our knowledge, there are no existing works on visualizing 
student progression patterns of an institution-wide student population.






\subsection{Technical Background}

In order to extract student flow patterns, relationships among academic entities such
as students, courses and majors are analyzed. The first relationship modeled is major similarity.
Over time, as the courses that students take become more specific, the
possibility of switching majors declines. This creates a hierarchy
of majors based on how similar they are. eCamp uses hierarchical clustering
to exploit this temporal hierarchy of majors.

Hierarchies have been visualized in many ways. Dendrograms and TreeMaps
\cite{johnson1990treemaps} are two traditional ways of visualizing hierarchical
structures. However, hierarchies with a temporal aspect cannot be adequately
represented by these traditional schemes. One alternative in such cases is using
Radial trees. Radial trees have been used in visualizing phylogenetic trees to
show how biological species evolve over time \cite{huson2007dendroscope,zhang2012evolview}. However, they are unable to
depict the proportion of elements that go through the hierarchy. 
Alternatively, sunburst graphs have been used to show how a population divides, going
from one level of a hierarchy to another. PathRings uses sunburst graphs to show
biological pathways \cite{zhu2015pathrings}. One important aspect of sunburst
graphs, however, is that they do not convey flow. 
Temporal flow is often visualized using Sankey diagrams
\cite{cullen2010efficient,alemasoom2015energyviz} or Sankey-like structures
\cite{xu2013visual,wu2014opinionflow}. Inspired by these techniques, eCamp uses
a variation of a radial tree that visualizes student progression with Sankey-like
edges to better convey the flow of students through time.

The second such relationship is the
correlation of student success, represented as grades between courses. 
The course-course correlation is measured by Pearson's coefficient.
The graph formed by all courses of each major is a standard graph, which we
render as a node-link diagram~\cite{reingold1981nodelink}, as typically done in the field.

\section{Driving Application}
\label{sec:driving_app}
\subsection{Overview}

\begin{figure*}[t]
  \centering
    \includegraphics[width=0.9\textwidth]{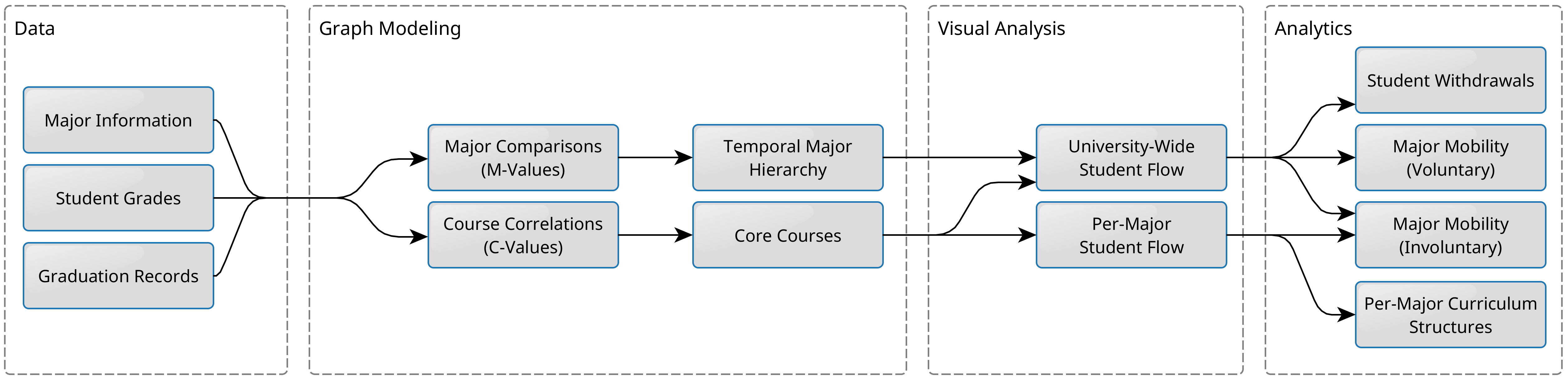}
    \caption{
    The concept overview, showing the process of transforming the data into
    analytics which can be used to discover patterns in student progression.
    From the three data categories (Section~\ref{sec:data}), there are two major
    channels of modeling (Section~\ref{sec:data_models}), two visualizations
    (Sections \ref{sec:node_link} \& \ref{sec:radial_tree}) and four classes of
    analytics (Section~\ref{sec:case_studies}).
    }
  \label{fig:concept_workflow}
\end{figure*}



Our aim in this work is to develop an analytical framework that shows how the
student population as a whole achieves its ultimate goal, graduation in a
major of each individual's choice, while exercising free will despite being
given a
choreographed script for its actions, the catalog. The outcome of the work
provides new insight into university administration,
and causes faculty and administrators to question
how the system has been designed and whether the effects seen in
the student population data are intended or not.

\subsection{The Data}
\label{sec:data}

Like most organizations, universities collect and maintain their data assets
independently by each campus office or department
for their own purposes.
Our data has come from multiple
campus offices, all of which maintain the data in the campus wide
ERP system called Banner. After processing, there are three main categories of data
as shown in Table~\ref{table:raw-data}.
These data include records from 144,798 students and over 400
majors over a period of 16 years.

\begin{table}[h]
	\centering
	\fontsize{8}{10}\selectfont 
	\caption{
	  eCamp Data Categories and Sizes
	}
	\bgroup
	\def\arraystretch{1.2}
    \begin{tabular}{| l | r | r |}
        \hline
        Category & Number of Entries & Size (MB) \\ \hline
        Graduation Records & 100,239 & 33 \\ \hline
        Student Grades & 4,723,835 & 461 \\ \hline
        Major Information & 436 & $<$1 \\ \hline
    \end{tabular}
    \egroup
  \label{table:raw-data}
\end{table}

Graduation records provide information on which major students
graduated in. Student grades provide information on when students took
courses and what their grades in those courses were. 

Major codes is a unique identifier of a major in the database. Major information provides the ability to connect major codes used in the graduation records with major names. The names are not necessarily unique. This can happen when a major gets revamped significantly and receives a new major code for instance. For analysis, we use the unique identifier. However, However, for user-friendly reasons, the visualizations still use major names as text labels.

\subsection{Analytics Needs}

To better understand the needs of faculty and administrators,
the authors have met with faculty, and administrators at the department level and
in the Provost's office.

The Provost's office pointed to a need for analytics to help them evaluate how advising
programs are affecting student retention and time-to-graduation, on a campus level,
college level, and specific to individual majors, especially those with a large student
population.
In particular, they noted that a current gap in knowledge across the nation
is identifying how success in general education courses may affect a student's success
in different majors. 
Furthermore, when a student transitions from one intended degree program
to another, what kind of transitional advising can be provided before, during and after the change?
These questions cannot be approached feasibly unless there are tools to reveal the underlying
patterns, missing links, and problem areas.

Similar to the campus level desire for improvements, department heads
need to better understand student progression patterns through their majors. They want to know
which courses play a central role in student
success, and whether the curriculum is working as designed  (e.g. are the gate keeping courses
serving their intended purpose). Furthermore, in relation to student progression,
departments want a better understanding of student success, retention, and diversity issues from freshman year to the senior year.

Based on these needs, we identified three classes of curricular
information to extract from the dataset:
(i) student flow through university wide curriculum and progress patterns (i.e. graduation vs. dropouts)
(ii) per-major curriculum structures,
and (iii) student flow through each major.

\section{Modeling Multi-Level Relationships}
\label{sec:data_models}


We summarize the overall architecture of eCamp in Figure~\ref{fig:concept_workflow}.
The steps to model major-major relationships and course-course relationships
are described in this section. After the models are constructed, the steps to
perform visual analysis of student flow are described in Section~\ref{sec:vis_analytics}.
The analytical tasks from the perspective of university administrators are in
Section~\ref{sec:feedback}.


As shown in Figure~\ref{fig:concept_workflow}, our data sources 
include information about three academic entities---majors, courses, and students. 
A major is comprised of both courses and students, and a course
is comprised of students.
%
%
With this population-scale student data, there are a variety of relationships that can be studied.

First, by analyzing how courses are shared among majors, data-driven relationships
among all of the majors that students can choose from can be found.
This knowledge on the major-major level depends on knowledge on the course-major
level, which in turn depends on the course-course level and fundamentally the
student-course level relationship.  This observation of the multi-level nature
of the relationships drove the design of eCamp's modeling, visualization and
analytics components.

Second, how courses relate to each other by the empirical
order in which students take these courses can be identified, and by the correlations
in grades achieved by the student cohort.
Previously, the only way to identify such course-course relationships
at scale was to resort to the pre-requisite or co-requisite relationships
defined in catalogs. However, it is desirable to observe if students progress through the catalog the way it was intended by the university.



\subsection{Major-Major Relationship}
\label{sec:student_major}
Many courses that students take through their studies are often shared between a set of majors. This means that majors have an overlapping relationship with one another. As students take more courses that are specific to their own major, it becomes less possible and less probable for them to switch majors. We calculate this overlapping relationship based on the student records from those who graduate in those majors and the set of courses that they have taken. 

%
%
%
%

\subsubsection{M-Value}

Calculating major-major relationships begins by estimating the degree to which students in a single major
will take a set of courses.
Given a major $A$ and a set of courses $C$, the estimate,
$M_{A}$, is

\begin{equation}
\label{eq:m_value}
M_{A} = \sum_{c_i \in C} \frac{s_{A}}{\left| S_{c_i} \right|_2 \left| A \right|}
\end{equation}

where $s_{A}$ is the number of students from major $A$ in the course $c_i$,
$|A|$ is the total student population of major $A$.
$S_{c_i} = \left[ s_{m_1},s_{m_2},...,s_{m_n} \right]$ is a vector of
counts of students in $c_i$ from all of the $n$ different majors.
$\left| S_{c_i} \right|_2$ is the Euclidean norm of the vector $S_{c_i}$ and is
computed as $\left| S_{c_i} \right|_2 = \sqrt{\sum_{k=1}^{n} s^2_{m_k}}$.

In Equation~\ref{eq:m_value},
$\frac{s_{A}}{|A|}$ corresponds to the
probability that students in major $A$ take course $c_i$.
The per-course scores are then tallied up across the whole set of courses
to form the overall M-Value for the major.

Some courses are taken by a much broader group of students than others.
For example, introductory English courses have
very little specificity in terms of majors, because they are shared
by the entire student population.
The additional term $\left| S_{c_i} \right|_2$ is introduced to reduce
the weight of those general courses.
This means that the final M-Value metric will be weighted towards courses
which are shared between small sets of majors.
A high M-Value means the given course set $Courses$ has a high
specificity to a major.
If the students in a major are not taking the courses in $Courses$,
the M-Value will be low.

The M-Value essentially measures the affinity between a major and a set of courses.
In other words, on the basis of a fixed set of courses, $C$, one can compute the affinity
measure of all of the majors on campus with that set of courses, $C$.
For example, if the course set $C$ consists entirely of bio-engineering courses,
the M-Values computed for each major can help to rank the similarities of all of the majors
on campus with bio-engineering.

\subsubsection{Major-Major Relationship Graph}
\label{sec:major-groups}

\begin{figure}[t!]
  \centering
    \includegraphics[width=.49\textwidth]{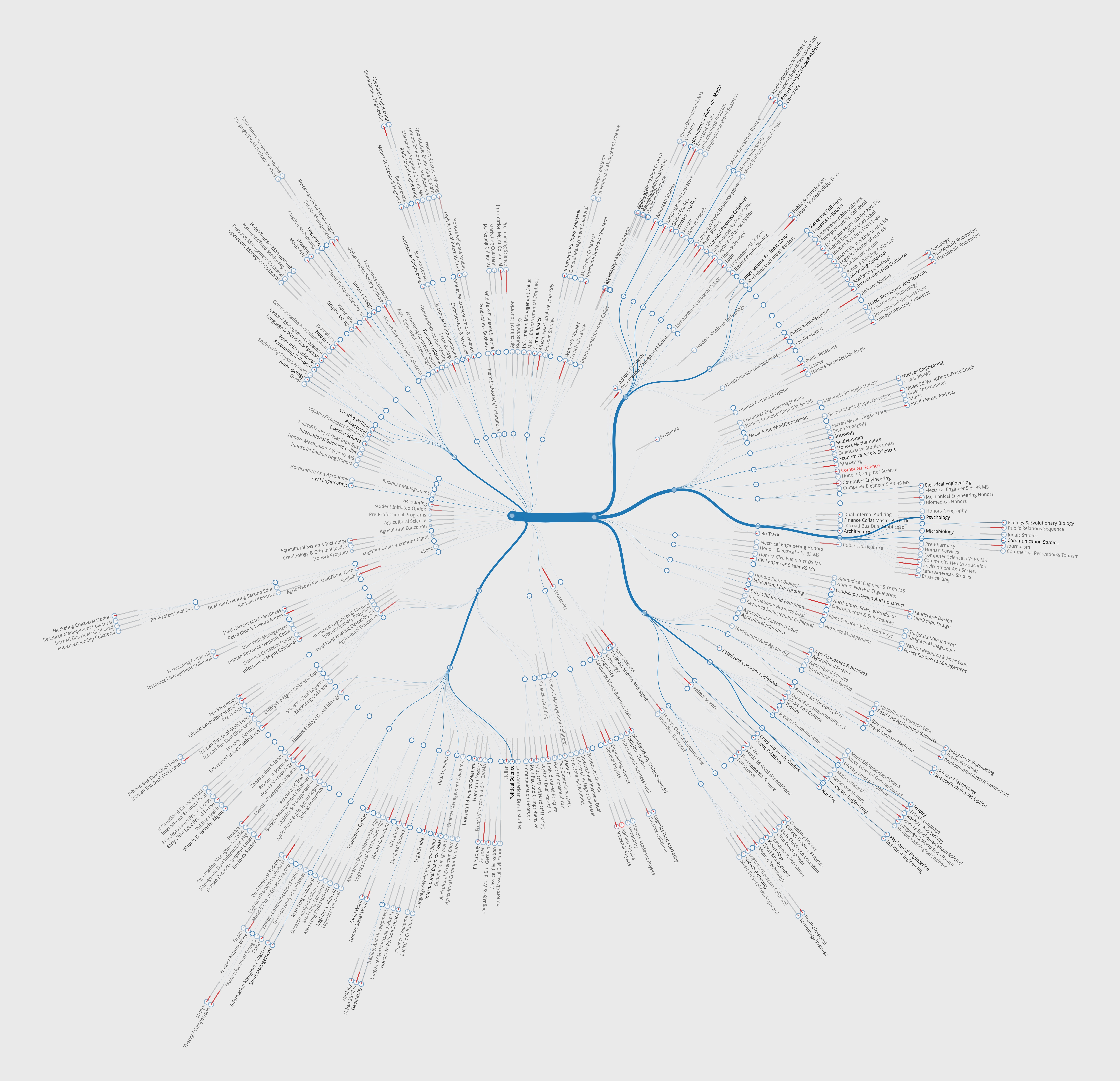}
    \caption{
    University-wide student flow visualization showing student
    progression semester by semester.
    Each leaf node corresponds to a single major,
    and the thickness of the paths corresponds to the number of students
    progressing through nodes. Computer science is the red major towards the
    right of the visualization.
    }
  \label{fig:radial_tree}
\end{figure}

Using the M-Value, we can capture the similarity between all majors on campus. This similarity for two majors, $A$ and $B$ is calculated as:

\begin{equation}
M_{A,B} = \frac{M'_{A} + M'_{B}}{2}
\end{equation}

where $M'_{A}$ is calculated for major A according
to Equation~\ref{eq:m_value}, but using
the course set taken only by students in major B.
$M’_{B}$ is calculated for major B, but using the course
set taken only by students in major A.

Suppose major A is computer science and major B is math.
Then $M'_{A}$ measures the affinity between the major of computer science and
the math major's courses.
$M'_{B}$ measures the affinity between the major of math and
the computer science major's courses.
$M_{A,B}$ is an average of those two metrics and is the same value as $M_{B,A}$.

One can now gain a more precise control of the model by controlling which set of
courses are used to compute the M-Values.
For example, one can make major-major
comparisons based on stages of a student's education,
by including in the course set, $C$, only those courses taken typically
by the student population during the corresponding stage
(such as freshman year vs. sophomore year or later).
The resulting major-major similarities computed using M-Values will then
vary from freshmen, sophomore, junior to senior year.




Conceptually, it is desired to then model and visualize majors gradually diverging from each other
as time progresses for the student population on a per-semester basis, and observe how students move or dropout along the way. 

The algorithm itself is a top-down clustering approach, beginning with all majors in a
single group, with the result being a tree. At each stage (i.e. academic semesters during
freshman, sophomore, junior, and senior year),
an M-Value similarity matrix using each semester's course set is calculated.

The tree is initialized to have only the root node with all majors belonging to it. 
Then, the process proceeds step by step.
Starting with the first semester of the freshman year -
the courses typically taken during that semester are chosen,
and a similarity matrix is produced.
In each step, one new level in the tree is created.

The process then proceeds to the next stage - the second semester
during freshman year.  The above process is recursively repeated, treating
the leaf nodes (first-semester division) as sets of majors to further
divide, selecting courses typically taken by that group of students as
the basis to determine how to make the division through clustering.
This process continues through all eight
semesters in a 4-year tenure of each student.

In each step, tree nodes are partitioned using a similarity matrix based on M-Values
that are computed from course sets specific to that semester.
In addition, when a tree node contains only one major, that tree
node is not further subdivided.


The hierarchy of majors that results from this algorithm has a very clear
interpretation. Figure \ref{fig:radial_tree} is a rendering of the hierarchy as a Sankey-like radial graph. Each leaf node corresponds to a single major, and each internal
node represents that the set of majors below it were considered to be similar
majors at that semester.

This hierarchy of majors also shows how earlier choices
made by students lead, or limit, them to certain majors as they
progress towards graduation. This effect of temporal bifurcation
cannot be captured through traditional methods.

\subsubsection{Student Dropout Patterns}
While junior and senior students usually have a ``declared" major, they can still change
their major without going to the registrar's office to update their records. 
In addition, although freshman and sophomore students may also have a ``declared major", many of them
are in an exploration stage of their studies and they may be taking preparation courses that can lead
to a few different majors. Effectively, their final major is unclear at that point. 

Both of these situations can cause significant data quality issues if we analyze
solely based on their ``declared" majors.
When it comes to analyzing for patterns of student dropouts, we need to make best-effort
estimates of a student's intended major based on the data available.

For this, we look at the courses that the students have taken and measure the amount of overlap between those courses and the courses of each potential major. The higher this overlap ratio is for a major, the more likely it is that they were pursuing that major. 

Counting the number of estimated dropouts for a major can introduce uncertainties. For example, the intention of a student that has only taken two courses are more unclear than a student that drops out after having taken ten courses. Another potential caveat with this approach is a scenario where a student has changed major without updating his/her major in the university records, and then dropped out. By the data, it is difficult to not count the dropout as the previous major. 

We do recognize these potential issues, and we account for this by showing the average overlap ratio for each major in a tooltip. The tooltip is shown when a user hovers over a major. 

In Table \ref{topmajors}, we show the top-5 majors in the database by number of graduates, and the average degree of overlap between courses taken by dropouts vs. the full curriculum of the best-matched major.
If the average overlap is high, then these are more likely to have intended to graduate from that major. If the average
shows low overlap, then very likely these are students dropping out early in their studies. This could hint that general education courses are causing the dropout rather than specialty courses of academic departments.

\begin{table*}[]
\fontsize{8}{10}\selectfont 
\centering
\caption{Top five majors by number of graduates. The average overlap shows the average percent of overlap between courses of students that dropped out and the courses of the major.}
\label{topmajors}
\bgroup
\def\arraystretch{1.2}
\begin{tabular}{|l|l|l|l|}
\hline
Major Name                                     & Number of Graduates & Estimated Number of Dropouts & Average Overlap \\ \hline
Psychology                                     & 3092                & 159                          & 63.71\%               \\ \hline
Political Science                              & 1263                & 52                           & 61.40\%               \\ \hline
Journalism \& Electronic Media                 & 1094                & 51                           & 72.79\%               \\ \hline
Communication Studies                          & 1012                & 59                           & 70.93\%               \\ \hline
Biochemistry \& Cellular and Molecular Biology & 826                 & 12                           & 75.70\%               \\ \hline
\end{tabular}
\egroup
\end{table*}

The total number of estimated dropouts for each major is shown in the major-major graph using a red and gray bar. The percentage of the red bar over the gray bar represents the dropout percentage.

\subsection{Course-Course Relationships}
\label{sec:course_major}

Diving into a more detailed view, we look at student success and student progression at a departmental level. In academic departments, student progression is typically represented by pre-requisite relationships in course catalogs. However, many courses do not have pre-requisites. Additionally, some pre-requisite rules are not always enforced. Therefore, the actual progression of students cannot be captured effectively. In our available data, we found that students grades are the closest variable that when combined with temporal information about courses, can represent progression and success in a major. Specifically, we quantify how the courses taken by students in a major are structured with respect to when courses are taken by the
students, as well as how courses are correlated in terms of student grades. With this knowledge, per-major curriculum structures can be determined. We believe other variables, such as instruction style, grading practices, rigor, etc. can help make the measure of student progression more accurate. However, these variables were not available. 

For this purpose, we first calculate course-to-course correlation
of student success. We then take these correlations and determine which courses are
most-highly correlated with all other courses and at what point in time each
course is being taken. 

\subsubsection{C-Value}

The approach for determining course correlations is the C-Value metric. The
C-Value, informally, measures the similarity between two sets of
grades, while accounting for the size of these sets. To begin the formal
discussion, the C-Value is heavily based upon the Pearson Correlation
Coefficient (PCC), which is commonly used for studying linear
correlation between variables.

Let $X$ be a collection of grades for course $A$, and $Y$ be the
collection of grades for course $B$. For these sample populations the PCC,
$r_{A,B}$, can be described as the sample covariance of $X$ and $Y$ divided by
the product of the sample variance of $X$ and the sample variance $Y$. This yields

\begin{equation}
r_{A,B} = \frac{\sum_{i=1}^{N^{A,B}}(X_i - \bar{X})(Y_i - \bar{Y})}{\sqrt{\sum_{i=1}^{N^{A,B}}(X_i - \bar{X})^2}\sqrt{\sum_{i=1}^{N^{A,B}}(Y_i - \bar{Y})^2}}
\end{equation}

where $\bar{X}$ and $\bar{Y}$ are the sample means for $X$ and $Y$,
respectively, $N_{A,B}$ is the number of students who took both course $A$ and
course $B$, and $X_i$, $Y_i$ are specific student grades.

By using PCC, one can see the correlation between courses based upon how
students performed within both of these courses. However, in this use case, 
PCC is insufficient without a final step. Consider a situation where only five
people took a course typically unrelated to a major and then all went on to do
well academically in the major. One might incorrectly determine that this course
is highly correlated with success in the major. To correct for this, the final
correlation metric, the C-Value, is scaled by $N_{A,B}$, producing

\begin{equation}
C_{A,B} = N_{A,B} \cdot r_{A,B}
\end{equation}

Using the pairwise C-Value, one can calculate correlations between all
courses within a major, and capture their similarity in terms of grades. 

\subsubsection{Per-Major Course-Course Relationship Graph}
\label{sec:per-major-graph}
\begin{figure*}[t]
    \centering
    \includegraphics[width=\textwidth]{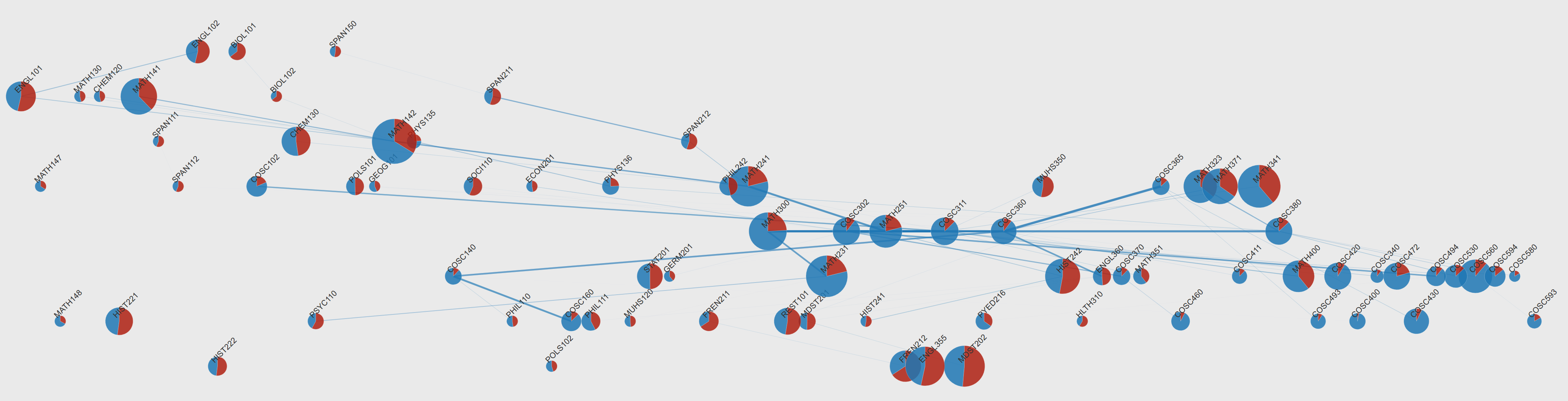}
    \caption{
    Per-major curriculum diagram. Thicker
    edges between courses signify higher correlation between student grades in
    these courses. The major shown is computer science. The ``core courses''
    are MATH 300, COSC 302, MATH 251, COSC 311, COSC 360, and COSC 380. The size of the nodes corresponds to to the percentage of students failing the class. Inside each node, the red portion corresponds to the ratio of female students, while the blue shows the ratio of male students in the class. 
    }
  \label{fig:node_link}
\end{figure*}
Calculating pairwise C-Value results in a similarity matrix, which can alternatively be thought of as an undirected, fully-connected, edge-weighted graph (shown in Figure \ref{fig:node_link}). In this graph, each node is a single course and each edge weight corresponds to the C-Value between two courses. 

The graph helps capture success progression through a major in two stages. First, courses that represent overall success in a major can be defined as those that correlate most with all other courses of the major. With this in mind, we can sort courses based on how representative they are of success in a particular major. We call the most representative courses, ``core courses''. Returning to the similarity matrix notion of the C-Value results, this is done by finding the rows/columns with the highest sum. Second, we can calculate where a course fits temporally in the real-world curriculum. This is done by determining the average time, or semester, during which students take the course. Looking at core courses and their correlations, administrators can see if in practice the courses exhibit the logical organization that they had intended for them. A visualization of this model is presented in Section \ref{sec:node_link}.

Additionally, in extracting these core courses, it becomes possible to fill a hole
in the data source---the major being pursued by students who withdrew from the
university is not known. Using each major's core courses and the set of courses that each student took, what
major the student was pursuing at the time of withdrawal can be predicted. This is owing to the fact that core courses are more representative of a major than other courses. The approach for
determining a student's intended major is to, for each major, measure the
percentage of core courses that the student took. The more core courses a
student takes, the narrower their options of switching majors would be.
Therefore, the higher that percentage is for a major, the more probable it
would have been for that student to graduate in that major. The intended majors
for student withdrawals are included in the university-wide student flow
visualization described in Section~\ref{sec:radial_tree}.

\section{Analytical Tasks}
\label{sec:vis_analytics}


The course-major and temporal hierarchy models described in
(Section~\ref{sec:data_models}) can be used to develop visualizations of student flow through the curriculum on a per-major basis and student flow through the temporal hierarchy of majors on a university-wide basis.
The university-wide student flow visualization (Section~\ref{sec:radial_tree}) employs the temporal major hierarchies and the per-major student flow visualization (Section~\ref{sec:node_link}) employs the course-major models.

Using these two visualizations, a series of results which
demonstrate the system's use for analytics purposes is presented in
Section~\ref{sec:case_studies}.

For ease of deployment and for enabling a crowd-sourced way of using a visual
knowledge discovery system, eCamp's user interface is fully web-based. The
web-based visualization is fully interactive and is created using D3.js and a
PHP backend. The data processing and modeling the multi-level relationships are
implemented in Python and take less than 5 minutes to compute on a regular
desktop.

\subsection{University-Wide Student Flow}
\label{sec:radial_tree}
Figure~\ref{fig:radial_tree} shows the university-wide student flow visualization.
For this, the primary aim is to show similarities between majors in terms of
how students progress towards graduation. Specifically, the goal is to show points
in time where students encounter critical decisions with regards to which
courses they take. At these critical decision points, the courses students
take may significantly limit their future options.

The visualization is based on the temporal hierarchy of majors
constructed in Section~\ref{sec:major-groups}.
The center of the visualization is the root of the tree of
majors. It is the starting point for all
students: before their first semester on campus when they have
the whole set of more than 400 majors from which to choose.

In this hierarchy, each leaf node is a major, and each level of the hierarchy
corresponds to a single semester. The paths show a flow of students from the
root to a leaf node.  Every step along the path reduces the student's choice of
potential majors, and eventually when reaching a leaf node, a student's
coursework will be virtually exclusive to major-specific courses for that major.
The width of the path corresponds to the size of the remaining student population
at logarithmic scale.

A secondary goal with this visualization is to show dropout patterns.
When a student drops out, the student's intended major can be predicted
using the method based on core courses as
described in Section~\ref{sec:per-major-graph}.
Each blue node on the path signifies that the subgroup has ``traveled''
together through the major hierarchy and reached a new milestone,
a new semester.  The grey and red line segments shown for each major represent 
the percentage of students who have dropped out of their intended major.
When the red line segment equals the gray line segment
in length, it means $100\%$ dropout. Correspondingly, when the red line
segment is half of the gray line's length, it means $50\%$ dropout.
When the user hovers over a major, a popup tooltip displays the actual number of students that have graduated or dropped out, as well as the confidence percentage for the dropout estimation. The opacity of the red dropout bars also reflects the confidence value. 


\subsection{Per-Major Student Flow}
\label{sec:node_link}

The per-major student flow visualization shows for each course which courses it
is strongly correlated with and when it is typically taken by students. An example of
this visualization with the computer science major at the university is shown
in Figure~\ref{fig:node_link}.

The primary design goal with this visualization is to cleanly show correlations
between courses. Accordingly, a node-link diagram was chosen for use, where
each node is a course and each link represents a grade correlation between
two courses. To allow users to quickly identify strong links, the
thickness of each link is scaled, where thicker lines correspond to higher C-Values.

Since there is a C-Value between all course pairs, this
has the potential to introduce significant visual clutter. To avoid this, the
user is allowed to specify a C-Value threshold, below which links between
courses are not shown.

The secondary design goals were to allow users to quickly identify ``core
courses'' and show when the courses were typically being taken. To incorporate
these goals, these two pieces of information are encoded into the courses'
horizontal and vertical positions in the diagram.

The horizontal position of a node indicates the average time at which the
course is taken by the student population. The vertical position of a course
represents whether a course is core or peripheral to the major, and is determined
by a course's total correlation to all other courses in the major. The ``core
courses'' are placed at the center of the diagram, with peripheral courses being
further away from the center. The number of ``core courses'' is a
user-controllable parameter. Since it is possible that nodes could occupy the
same position with this approach, a spring-force approach is used to force nodes
occupying the same position to have sufficient spacing. 

Additionally, each node provides gender information as a pie chart, and the size of each node represents the normalized percentage of failures in that course. For example, Figure \ref{fig:node_link} shows that math courses tend to be the bottleneck for students. Auxiliary information such as the exact number of students taking the course, and grade distributions are shown to users as they click on the nodes. 

Using this visualization, how students are moving through a major's
curriculum can be seen and whether the student population is progressing through
the major's curriculum in the way the faculty intended can be determined.

\subsection{Analytics Results}
\label{sec:case_studies}
With these two visualizations, eCamp is able to provide insight into the issues
raised by the users. The following subsections present results showing
how eCamp can enable users to form hypotheses regarding these issues.

\begin{figure}[h]
    \centering
    \includegraphics[width=.49\textwidth]{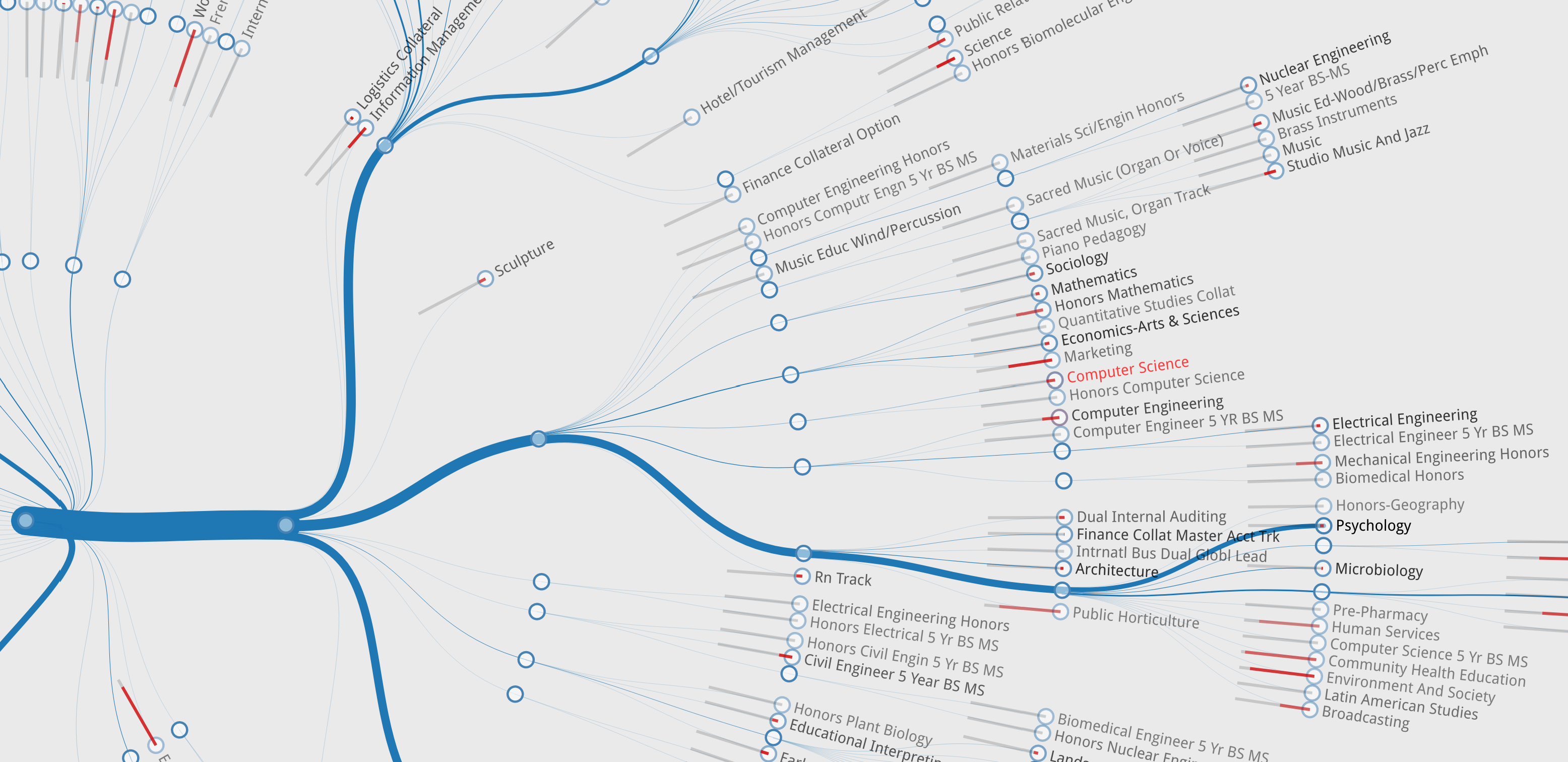}
    \caption{This branch of the radial graph contains the 
    university's computer and electrical engineering majors. These majors split apart from most
    other majors by the end of the 1st semester.}
  \label{fig:engineering_branch}
\end{figure}

\begin{figure*}[ht!]
    \centering
    \includegraphics[width=\textwidth]{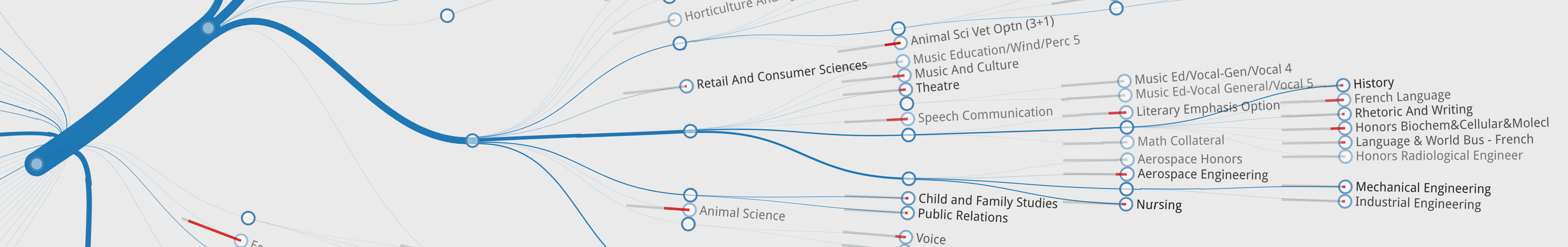}
    \caption{This branch of the radial graph shows the path towards mechanical engineering and industrial engineering. Students have until the fifth semester to choose between the two. }
  \label{fig:non_engineering_branches}
\end{figure*}

\subsubsection{Major Exploration Advising} 
Students who wish to explore different options before choosing a major must be made aware of how the courses
they choose to take limit their options of which majors they may pursue.
Figure~\ref{fig:engineering_branch} shows that students who are potentially
interested in electrical engineering have a very limited time in which to commit to it,
or they risk delaying their graduation. On the other hand, Figure~\ref{fig:non_engineering_branches}
shows that students have five semesters to choose between industrial engineering and mechanical engineering, if they want to graduate on time. This calls into question the university's policy of requiring a student to declare a major after 45 credit hours, as a student's interests may affect how long they have to commit to a major.

\subsubsection{Major Mobility Advising} Another common advising task is to help
students who wish to change majors. Consider a third-semester computer engineering
student who comes to the advisor and expresses a desire to change majors due to
a lack of interest in continuing computer engineering. Rather than
simply relying upon experience, the advisor uses the radial tree to determine
which majors would be a good fit for the student.

First, the advisor locates the path from the root node to the node
of depth 3 that contains computer engineering. This path can be seen in
Figure~\ref{fig:engineering_branch}. Then, the advisor records each of the child
majors from this node, and presents them to the student. Since at this point none of
these majors exhibit highly specified coursework, the student should
have little difficulty switching to any of them.

What if the student had not lost interest in computer engineering,
but instead failed the physics course in the second semester? In this case, it is likely in the student's best
interests to change majors. While the temporal major hierarchy can
again be used to determine which majors would be a good fit for the
student, the advisor must now also consider that the major
being switched to shouldn't have the same physics course as an
core course. Using the per-major curriculum diagrams for each of
the majors identified, the advisor notices that in the sociology major
the physics course is not close to the core courses, and recommends
to the student that he or she consider switching to it. Alternatively,
the advisor could use the ability to control the color saturation to show
the similarity of all majors to computer engineering. Majors which
exhibit a high similarity with it, but that are in different branches of
the tree, such as computer science, are good recommendations to the student.

\subsubsection{Course Correlations} 
\begin{figure}[h]
    \centering
    \includegraphics[width=.30\textwidth]{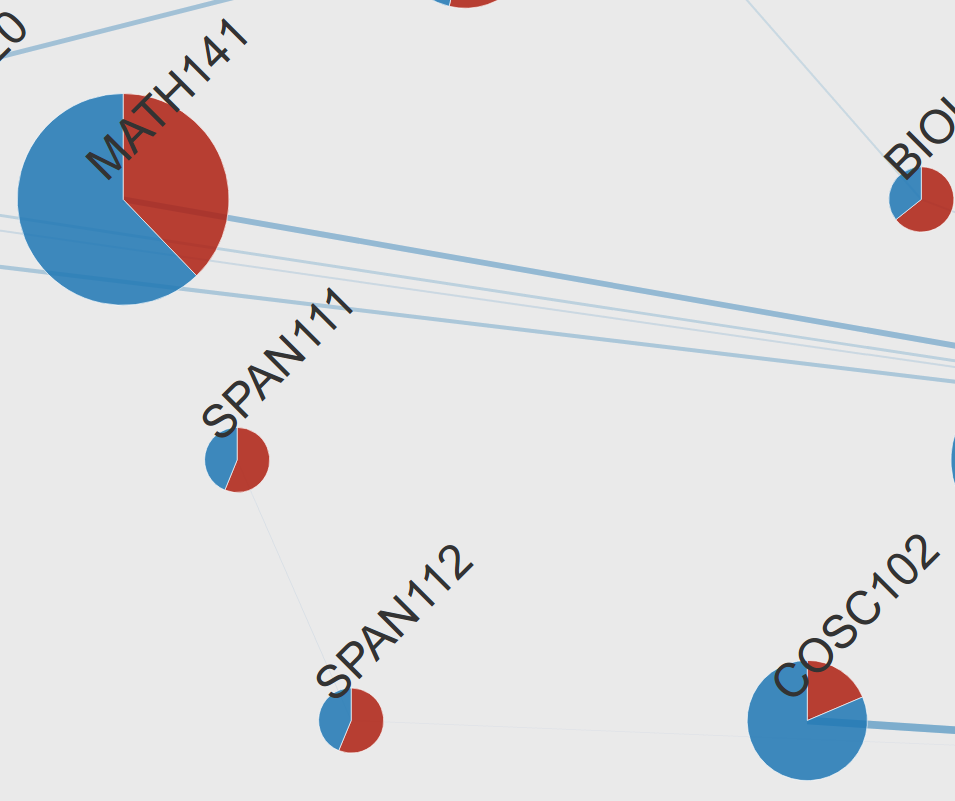}
    \caption{The presence and width of an edge represents the level of correlation in student grades between two courses. For example, because there is only a very faint edge between Calculus 1 (MATH 141) and the introductory computer science course (COSC 102), there is little correlation between them in terms of student success. }
  \label{fig:calc_correlation}
\end{figure}

eCamp's ability to show course correlations
has led to discovering a surprising result in the curriculum.
Figure~\ref{fig:calc_correlation} shows that there is no correlation between
Calculus I (MATH 141) and the introductory programming course (COSC 102) at
the university. For a long time students have been told to take these courses concurrently and
recently the faculty has formalized this advice by making them co-requisites.
However, seeing the lack of correlation between the two courses
calls into question whether or not this should remain the case (actually there is
a very weak {\em negative} correlation between the two courses).

\begin{figure}[t]
    \centering
    \includegraphics[width=.45\textwidth]{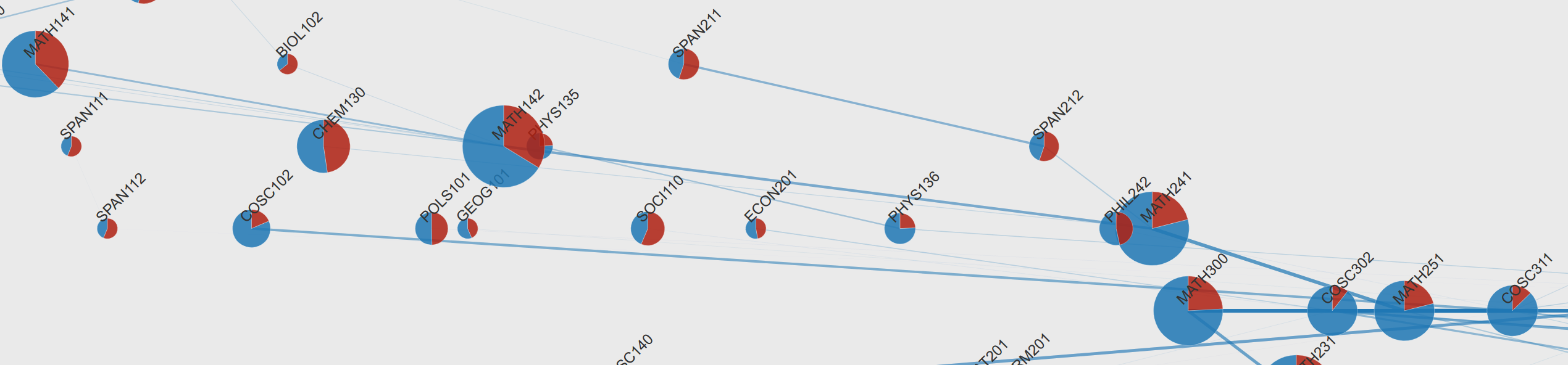}
    \caption{Course correlations between the Calculus sequence (MATH 141, 142, 241)
    and the core courses. The core courses can be spotted by the straight line and high correlation between them (bottom-left of the image). 
    }
  \label{fig:calculus}
\end{figure}

Despite this surprising result, it is seen in Figure~\ref{fig:calculus} that the Calculus
sequence as a whole does demonstrate correlation with two of the computer science
core courses, Linear Algebra (MATH 251) and Discrete Mathematics (COSC 311).
When seeing general education sequences that affect student success in a
major, the authors feel that it presents a good opportunity to encourage collaborations
between the major-level and university-level student support infrastructures.
As noted in the introduction, whether these courses should be core courses
in the CS curriculum is an interesting retention question, since they do not
correlate with student success in non-theoretical CS courses, and many CS graduates
will not engage in tasks requiring theoretical CS knowledge in their eventual jobs.

\section{Domain Expert Feedback}
\label{sec:feedback}
In this section, more detailed observations that
two of our domain experts derived from eCamp, as well as
feedback about potential improvements are presented.
The first was our department head, who is
intimately familiar with the per-major curricula for the Electrical
Engineering major. The second is a faculty member who formerly
served as Vice Provost of the university, whose priorities
are improving student retention and time-to-graduation. Both began
to ask questions regarding both the per-major and university-wide
curriculum that they had not previously considered.

Our department head examined the node-link diagram for
Electrical Engineering, and saw both patterns that
he had expected and patterns that he had not.
One finding was that some of the courses showing strong
grade correlations had the same instructor. He expressed a
preference that course success be independent of instructors and
instead be driven by the course's material. Additionally, he
expressed surprise that the courses meant to serve as gatekeeping
courses for Electrical Engineering did not show strong correlation
with success in most of the remaining curriculum, which led him
to wonder why that was the case.

The former Vice Provost felt that the node-link diagrams are
useful for evaluating majors in terms of how welcoming they are
to students switching to them. For example, she mentioned that
Classical Civilization is typically considered a found major, where
students who graduate with this major did not enter the university
planning on doing so. In such a major, the core courses would
ideally be very late in the student curriculum, which was the case
for Classical Civilization. On the other hand, engineering departments
would prefer for students to commit very early, so it would be
best for their core courses to be much earlier in the curriculum.

She thought that the radial graph was interesting to university-wide
administrators as it showed major branches of study available in
the university. Specifically, she saw four main arms of majors
and noted that the number of students graduating from each of
these arms was very uneven. This led her to ask questions regarding
how the university is distributing resources, and whether or not
this distribution matched the goals of the university.

She also felt that first-year advisors' work would benefit
from access to the radial graph visualization. Specifically, she saw
that the Journalism and Communications majors maintained shared curricula until very late in the student
career. This means that students are likely to choose one of these
majors before they have taken courses which would help them
determine which major is best suited to their interests. Hence,
it is important for first-year advisors to make sure that
students are informed that they should keep an open mind about which major
they like the most, as it should still remain possible to switch between
these majors very late into the curriculum.

Regarding improvements to eCamp, the former Vice Provost noted
that it would be useful to see how flow differs between
students with different financial backgrounds. She noted that
students from low-income backgrounds are considered to be at higher
risk of not graduating, and it would be interesting to be able to
see where these students are typically struggling. Additionally,
she suggested building a data science tool to predict when students are changing
majors and what majors they are changing to, as this could hint at why so many students take longer than 4 years
to graduate.

\section{Conclusion}
\label{sec:conclusion}
This paper has taken a data science approach to integrate and make sense
of previously disparate electronic student records, using
a framework that models relationships that span multiple
levels of entities: students, courses and majors. The prototype system, eCamp,
enables university personnel to leverage the information hidden in these
datasets to form questions and hypotheses about their curricula.
eCamp has been made available to several administrators at the authors'
university, and a selection of analytical questions that eCamp helped to raise
about student progression and retention are presented. In future works, we'd like to incorporate new data sources (e.g. student financial information) into our analysis.

\acknowledgments{
The authors would like to thank the anonymous reviewers of this and previous versions of the manuscript for their valuable comments and suggestions. The authors are supported in part by NSF Awards OCI-0906324, CNS-1629890, and the Engineering Research Center Program of the National Science Foundation and the Department of Energy under NSF Award Number EEC-1041877.
}

\bibliographystyle{abbrv-doi}

\bibliography{ecamp_paper}
\end{document}